\newcommand{\CLASSINPUTtoptextmargin}{0.75 in} 
\newcommand{\CLASSINPUTbottomtextmargin}{20 mm} 
\documentclass[conference]{IEEEtran}
\IEEEoverridecommandlockouts
\usepackage{cite}
\usepackage{amsmath,amssymb,amsfonts}
\usepackage{algorithmic}
\usepackage{graphicx}
\usepackage{textcomp}
\usepackage{xcolor}
\usepackage{tikz}
\usepackage{subcaption}
\def\BibTeX{{\rm B\kern-.05em{\sc i\kern-.025em b}\kern-.08em
    T\kern-.1667em\lower.7ex\hbox{E}\kern-.125emX}}

\newcommand\copyrighttext{%
  \footnotesize \textcopyright 2020 IEEE.  Personal use of this material is permitted.  Permission from IEEE must be obtained for all other uses, in any current or future media, including reprinting/republishing this material for advertising or promotional purposes, creating new collective works, for resale or redistribution to servers or lists, or reuse of any copyrighted component of this work in other works.}
\newcommand\copyrightnotice{%
\begin{tikzpicture}[remember picture,overlay]
\node[anchor=south,yshift=10pt] at (current page.south) {\fbox{\parbox{\dimexpr\textwidth-\fboxsep-\fboxrule\relax}{\copyrighttext}}};
\end{tikzpicture}%
}

\begin{document}

\title{ \vspace{-0.5cm}
Just Noticeable Difference for Machines to Generate Adversarial Images
{}
\vspace{-0.4cm}
}

\author{
\IEEEauthorblockN{
Adil Kaan Akan,
Mehmet Ali Genc and
Fatos T. Yarman Vural}

\IEEEauthorblockA{Department of Computer Engineering, 
Middle East Technical University}

 \IEEEauthorblockN{\texttt{\{kaan.akan, mehmet.genc, vural\}@ceng.metu.edu.tr} \vspace{-1cm}}

}

\maketitle

\copyrightnotice
\begin{abstract}
 One way of designing a robust machine learning algorithm is to generate authentic adversarial images which  can trick the  algorithms as much as possible.   In this study, we propose a new method to generate adversarial images which are very similar to true images, yet, these images are discriminated from the original ones and are assigned into another category by the model. The proposed method is based on a popular concept of experimental psychology, called, Just Noticeable Difference. We define Just Noticeable Difference for a  machine learning model and generate a least perceptible difference for adversarial images which can trick a model.
The suggested model iteratively distorts a true image by  gradient descent method until the machine learning algorithm outputs a false label. Deep Neural Networks are trained for object detection and classification tasks. The cost function includes regularization terms to  generate just noticeably different adversarial images which can be detected by the model. The adversarial images generated in this study looks more natural compared to the output of state of the art adversarial image generators.

\end{abstract}

\begin{IEEEkeywords}
Adversarial Image,  Adversarial Attack, CNN, Regularization, Just Noticeable Difference.
\end{IEEEkeywords}

\section{Introduction}


Recently, systems using Convolutional Neural Networks (CNN) have emerged in many cutting-edge products. Unfortunately, most of the CNN based image classifiers or object detectors are not  secure enough against deceptions. In order to prevent these deceptions, generating as many deceiving images as possible is very crucial. 

In this paper, we suggest a method to generate authentic adversarial images, which successfully trick a machine learning algorithm. The suggested method is inspired from the concept of Just Noticeable Difference (JND). We adopt JND to machine perception by defining the least amount of change in an image to cheat a machine learning algorithm. After we introduce the formal definition of JND for machines, we gradually distort the image by a CNN to detect the least perceptible difference.
Basically, the suggested method selects an image from a  category and progressively updates it using a gradient descent based optimization algorithm  until the CNN model detects the change and assigns it to another category. The  resulting adversarial image can be considered as the just noticeably different image from the machine perspective. \par

The main contribution of this paper is to formalize the concept of  Just Noticeable difference by a set of  regularization techniques to generate adversarial images that look authentic and relatively noise free. The generated adversarial images not only cheat the machines, but also they can cheat humans, thus, look quite natural compared to the outputs of popular adversarial image generators. This study, also, points out that for a machine learning system, the concept of JND  is quite different than human perception. \par

\section{Related Work}

In most cases, humans cannot recognize the changes in the adversarial example, but the classifier misclassifies it.   Kurakin et. al., in \cite{b1}  state that even in the physical world scenarios, adversarial examples are still a threat to classifiers. They show this problem by using adversarial examples, which are obtained from a cell phone camera. They feed these adversarial examples to an ImageNet pre-trained Inception classifier and measure the accuracy of the classifier. They observe that most of the adversarial examples are misclassified.

A recent study of Sheikh et. al.  found some intriguing properties of neural networks \cite{b7}. For example, the adversarial examples generated on  ImageNet  were so similar to the original ones, even the human eye failed to distinguish them. Interestingly, some of the adversarial examples, also, got misclassified by other classifiers that had different architectures or they are even trained on different subsets of the training data. These findings sadly suggest that  deep learning classifiers, even the ones that obtain superior performances on the test set, are not actually learning the true underlying patterns that determine the correct output label. Instead, these algorithms have built a model that works well on naturally occurring data, but fail miserably for the data that do not have a high probability in the data distribution.

In  \cite{b5}, Papernot et. al. states that all existing adversarial example attacks require knowledge of either the model internals or its training data. They introduce a practical method of an attacker, controlling a remotely hosted DNN. The only capability of the  black-box adversary generator is to observe labels given by the DNN to chosen inputs. 

\section{Just Noticeable Difference for Machine Perception}

In experimental psychology, Just Noticeable Difference (JND) is defined as the least amount of change on some sensory stimuli in order to detect the change. In this study, we adopt this concept to machine perception, where machine detects the change on the input image and decides  that this is different from the original image. 

Let us start by defining the concept of Just Noticeable Difference (JND) for a Machine Learning system. This is a  critical step to generate an adversarial image from a real image,  which confounds a machine learning algorithm. The adversarial image with the least perceptible difference, in which the network  discriminates the true image from the adversarial image is called just noticeably different adversarial image.

Formally speaking, Just Noticeably Different adversarial image can be defined as follows: Suppose that  a  machine learning system $M$, generates a true label $y$ for  image $x$, i.e. $M(x) = y$. Suppose, also, that  image  $x$ is distorted gradually by adding an incremental noise $n(k)$ to generate an image $x(k+1)= x(k)+n(k)$ at  iteration $k+1$, where $x=x(0)$. Assure that  $M(x(k)) = y$ for all k= 1,...., K-1 and $M(x(k)) \neq  y $ for k $ \ge K $. $x(K)$ is called the Just Noticeably Different (JND) adversarial image.

Note that $K$ is the smallest number of iterations, where the machine notices the difference between the original image and the adversarial image. Note also that, the image generated at iteration K, is the least detectable difference from the true image $x$ by M, because; for all the generated images, $x(k)$ for $k\le K-1$,   $M(x(k)) = y$. The rest of the images $x(k)$  for  k $ \ge K $  are adversarial, i.e.  $M(x(k)) \neq  y $. 

\section{Generating the Adversarial image by Just Noticeable Difference}

In order to generate an adversarial image with  Just Noticeable Difference, $JND(M)$, for a machine $M$, we adopt the CNN architecture, suggested by  Gatys et. al. \cite{b6} for style detection. In their approach, they start from a random noise image and update it, so that the style of the image is captured. In our suggested method, we start from an original image $x$, which can be correctly classified by a model $M$ and update it  until the network notices the difference between the original image and the distorted image, in other words, we distort $x(k)$ until  $M(x(k)) \neq  y $. 

The input image $x(k)$ at each iteration is distorted by the gradient descent method to generate a slightly more noisy image, so that the model output can no longer generate the true label. 


We generate just noticeably different adversarial images for two different machine learning  tasks. The first one is image classification and the second one is  object detection task. In the  image classification task, we tried to correctly classify the image as we distort it. For the object detection task,   we used two approaches. In the first approach, we tried to match the output of the model to the bounding box coordinates and object class. In the second approach, we just matched the object class in a bounding box. 

\subsection{Cost Function }

The most crucial part of the suggested method is defining a cost function, which  assures that the generated adversarial image is as close as possible to the true image, while minimizing the loss function between the true label and assigned label. The cost function we propose has three regularization terms in addition to the loss function between the true label and the label generated by the model, as defined below:
\vspace{-0.25   cm}

\begin{align}
    Cost(x(k)) = Loss(\hat{y}, y) + \lambda ||x(k) - x(0) ||^2 \nonumber + \\ BR(x(k)) + TV(x(k)),
\end{align}

where $x=x(0)$ is the input image, $x(k)$ is the updated adversarial image at the iteration k, $\hat{y}$ is the output of the model given the updated adversarial image, $x(k)$, $y$ is the true output of the model. The first term, $Loss$ is the loss function depending on the model we are tricking. For example, if the task is image classification, we can use cross-entropy loss.  By adding three regularization terms to the loss term, we try to capture the Just Noticeable Difference for the model.  The first regularization term which measures the distance between the original image and the updated image, ensures that the updated image  is as close as possible to  the original image. The other two terms which are  Bounded Range loss, $BR$, and  Total Variation loss,  $TV$,  ensures the natural appearance of the images as will be explained in the next subsections. 

After we calculate the cost from  equation 1, we update the input image by using gradient descent. We use the following formula for update the adversarial image. 

\begin{equation}
    x(k+1) \leftarrow x(k) - \alpha \frac{\delta Cost(x(k))}{\delta x(k)} 
\end{equation}

where $x(k)$ is the updated adversarial image at iteration $k$, $\alpha$ is the learning rate, $Cost$ is the function defined in the equation 1. Note that the second term in the above equation corresponds to the  noise $n(k)$ added to the original image at each iteration:
\vspace{-0.1cm}

\begin{equation}
    n(k) = - \alpha \frac{\delta Cost(x(k))}{\delta x(k)} .
\end{equation}

Just Noticeable Difference for a model $M$ is, then, defined as,
\begin{equation}
  JND (M) =|| x(0)-x(K)||/x(0) ,
\end{equation}
 where the model detects the incremental change in the generated image $x(K)$ for the first time, at iteration $K$  and outputs a false label.  



\subsection{Regularization Techniques}
\label{regularization_section}


\label{sec:regular}
The second term of the cost of function, $L_2$ distance between the adversarial image and the original image is added so that the updated image is as close as possible to the original image. We use $L_2$ regularization term in all experiments because without using it, the adversarial image diverges from the original image.
The third and forth terms of the cost function, namely, bounded range,  and total variation  is added to assure the natural appearance of the generated adversarial images.   Mahedran. et. al \cite{b2} showed that  these techniques make a random image look more natural. Bounded Range loss simply penalizes pixels with higher intensities than maximum intensity and lower intensities than minimum intensity. Therefore, image stays in the range of 0 to 255. Total Variation loss makes image pixels' intensities same as their neighbors so that some surfaces become visible. 

\subsubsection{Bounded Range} 
In this regularization technique, a pixel is penalized if it has an intensity which is not in the natural image intensity range. We use the below formula for each pixel of the image. A natural image must consist of pixels with intensities with minimum 0 and maximum 255. Therefore, we penalizes each pixels which are not obeying this constraint. 
\vspace{-0.3cm}

\begin{align}
    BR(p) = \begin{cases} 
      -p & p \le 0 \\
      p - 255 & p \ge 255 \\
      0 & otherwise 
   \end{cases}
\end{align}

where $p$ is intensity of the pixel. This function is applied to each pixel of the image to assure that all the pixel values of the image is in the range of [0 - 255].

\subsubsection{Total Variation} 
In  natural images, a pixel has a "similar" intensity value with its neighbors. This property can be partially simulated by  total variation regularization technique. In this technique, a pixel is penalized if it does not have the "similar" intensity with its neighbors. Therefore, we penalize the image to force all pixels to  have similar intensity values with their neighbors. We use the below formula to penalize the variation among pixels;
\vspace{-0.25cm}

\begin{align}
    TV = \frac{1}{HW} \sum\limits_{uvk} [(x(v,u + 1,k) - \nonumber
    x(v,u,k))^2 + \\ (x(v+1, u , k) - x(v, u, k) )^2] ,
\end{align}

where H and W are  height and width respectively, u,v,k are values to iterate over the dimensions height, width, and depth. \\

\vspace{-0.25cm}

\section{Experimental Results}

\begin{figure}
\centering
\vspace{-0.7cm}
\begin{subfigure}[b]{0.5\textwidth}
   \includegraphics[width=1\linewidth, height=2.5cm]{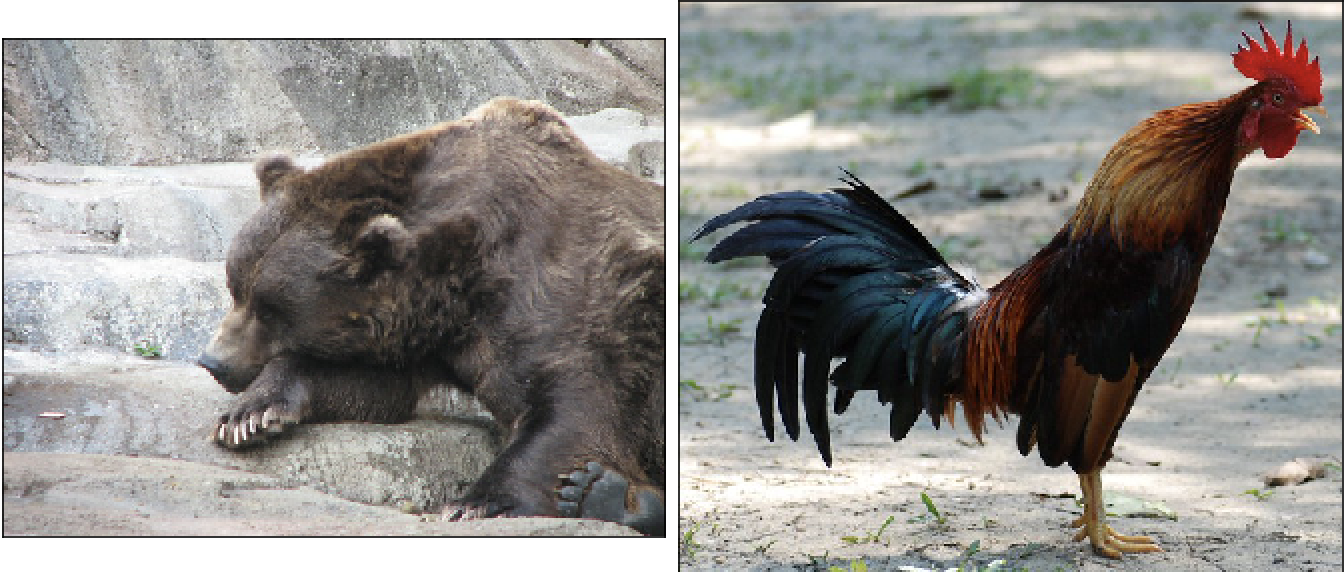}
\end{subfigure}

\begin{subfigure}[b]{0.5\textwidth}
   \includegraphics[width=1\linewidth, height=2.5cm]{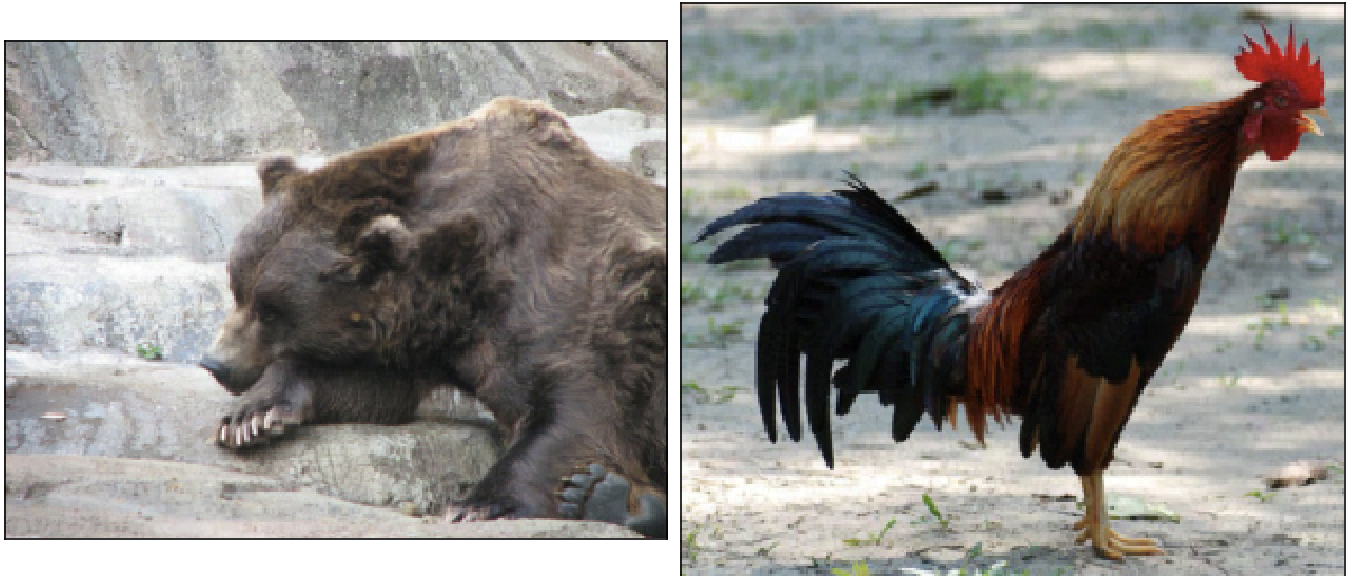}
\end{subfigure}

\begin{subfigure}[b]{0.5\textwidth}
   \includegraphics[width=1\linewidth, height=2.5cm]{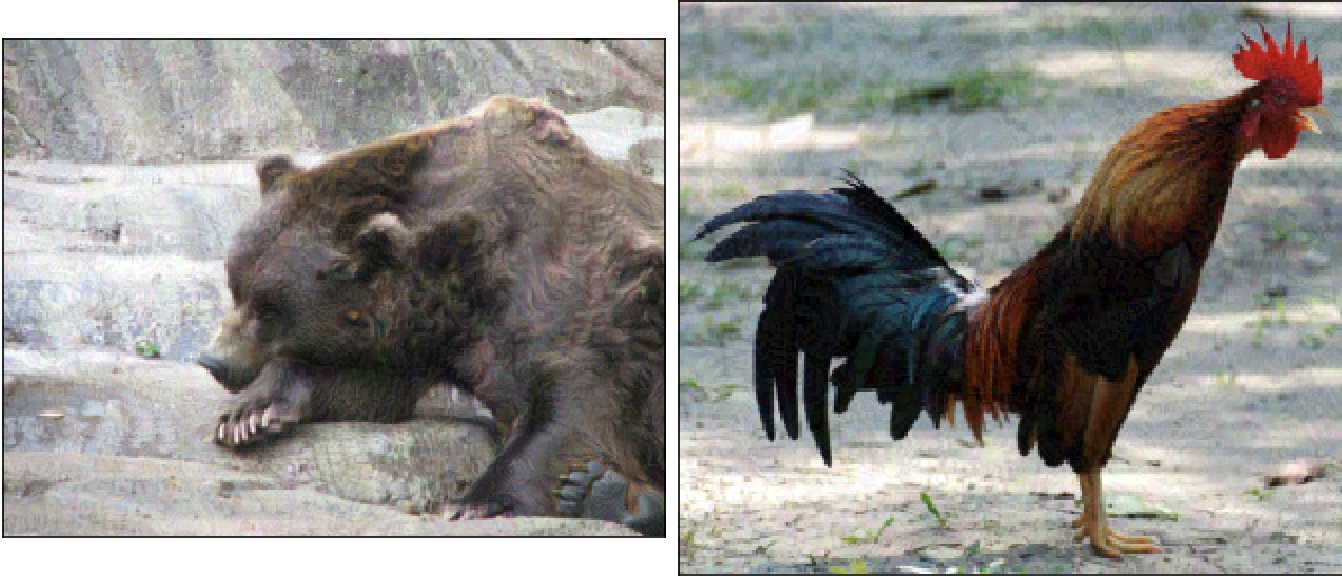}
   \label{fig:Ng2}
\end{subfigure}
\vspace{-0.5cm}
\begin{subfigure}[b]{0.5\textwidth}
   \includegraphics[width=1\linewidth, height=2.5cm]{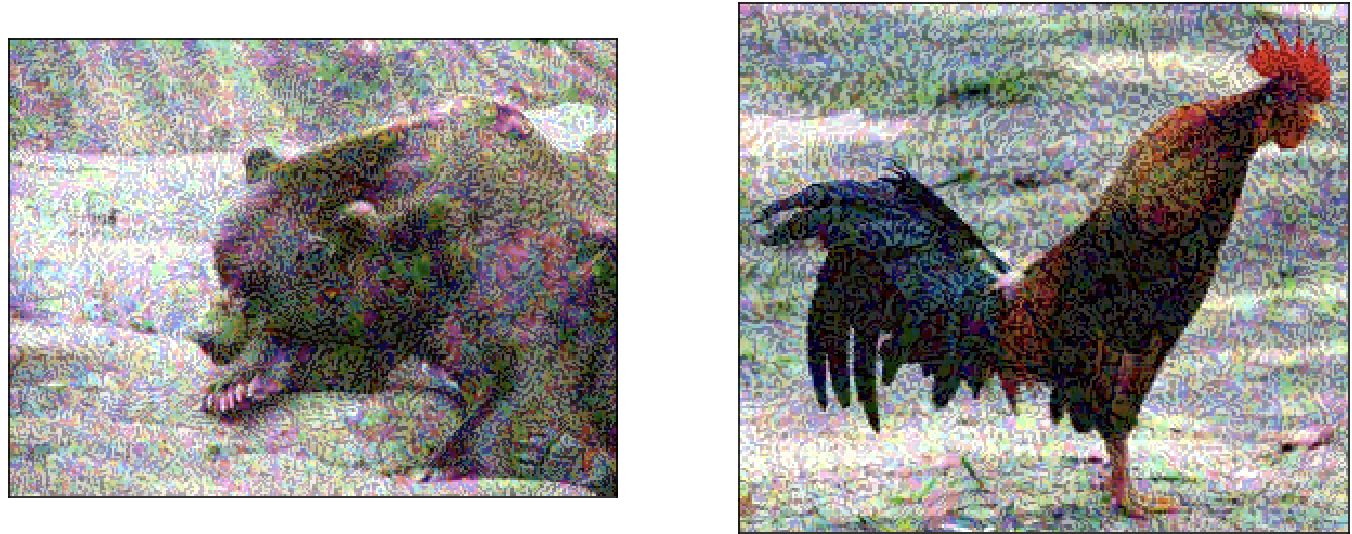}
   \label{fig:Ng2}
\end{subfigure}
\caption{Experiments in ImageNet dataset: First row: Samples from original images. Second row: Just Noticeably Different Adversarial Images with using regularization functions at the time when they first trick the network. Third row: Samples generated by regularization functions at the end of the generation. Bottom row: Samples generated without any regularization. All the samples trick  classifiers successfully.}
\label{fig:imagenet_images}
\vspace{-0.1cm}
\end{figure}

\begin{figure}
    \centering
    \includegraphics[width=0.5\textwidth]{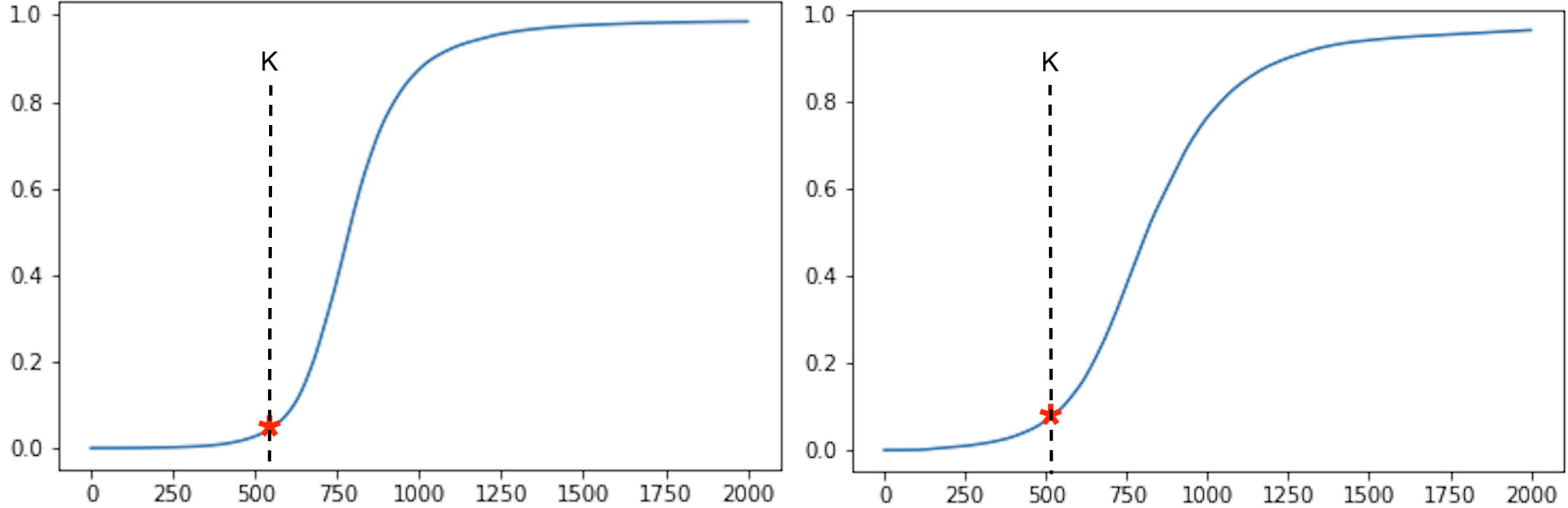}
    \caption{The confidence scores of the updated bear and rooster images, after K=509 and 561, respectively, the images  successfully trick the network. Therefore, Just Noticeably Different Adversarial image is generated at K=509 and 561, respectively. Y-axis shows the confidence score, in 0-1 scale and X-axis shows the number of iterations}
    \label{fig:conf_scores}
    \vspace{-0.4cm}
\end{figure}

\begin{figure*}[h]
    \centering
    \includegraphics[width = \linewidth, height=3cm]{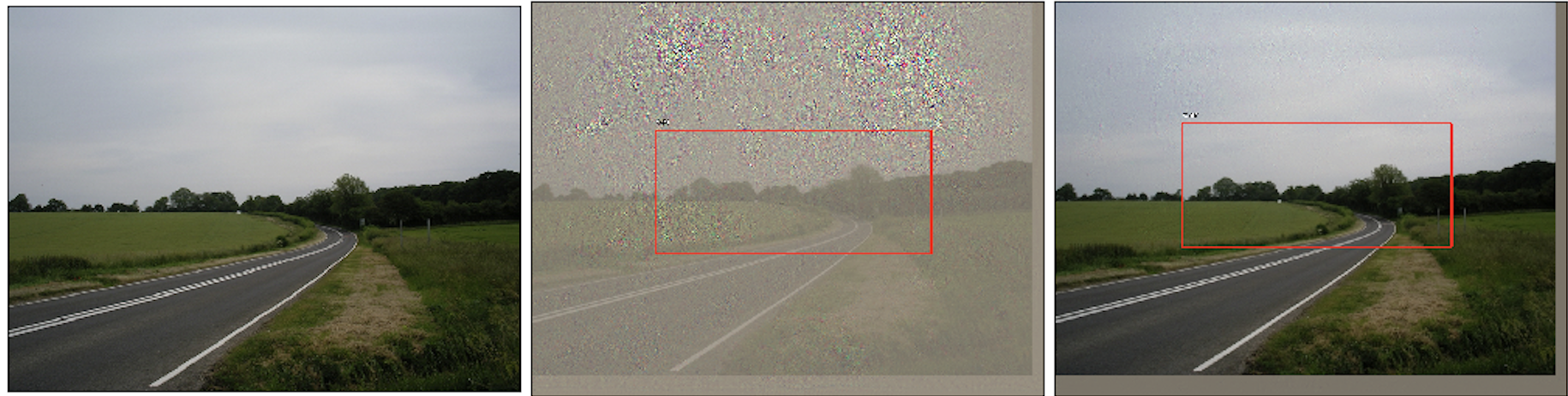}
    \caption{Generated images to trick the object detector. On the left, original image, in the middle, generated image without regularization, on the right, generated image with regularization, i.e., Just Noticeably Different Adversarial Image.}
    \label{fig:obj_detect}
    \vspace{-0.35 cm}
\end{figure*}

In this section, we demonstrate that the suggested method generates  adversarial images, which are indistinguishable to  human eye in high resolution images,  are capable of tricking the models. We conduct  experiments in ImageNet, CIFAR10 and MS COCO data sets in two different machine learning tasks, namely image classification and object detection. First, we generate JND adversarial images to trick a pre-trained Inception v3 image classifier  on  ImageNet and  CIFAR10. Second, we generate JND adversarial images to trick ResNet object detector. In order to  examine the power and weaknesses of the suggested  method, we illustrate some of  the generated images. We, also, measured some image quality  metrics to make a numerical comparisons between   the original images and the generated images. 

The results indicate that using the suggested regularization techniques in a relatively high resolution dataset, such as ImageNet, we observe a substantial  increase in the naturalness of the adversarial  image. However,  in a relatively low resolution dataset, such as CIFAR10, regularizations  does not significantly improve the quality of generated images. We, also, verify this observation by image quality metrics \cite{b8, b11, b10, b9}. We use pre-trained Inception v3 model \cite{b12} for classification experiments, and RetinaNet \cite{b3} trained on MS COCO \cite{b4} for object detection experiments.

\vspace{-0.1cm}

\subsection{Classification Experiments on ImageNet}

We generate 3 different sets of adversarial images which can successfully trick image classifiers. The first set consists of  the Just Noticeably Different adversarial image generated with all the regularization terms. This is the image set  that tricks the network for the smallest  iteration number K. The second set of  adversarial images are generated at the final step of the iterations. The third set of adversarial images are generated with no regularization. 

Note that, using regularization significantly increases the naturalness of the generated images. This result shows that regularization is very effective in high resolution images  (at least $224x224$). The generated images can be seen in Figure \ref{fig:imagenet_images}. While increasing the quality of  images, using regularization does not decrease the confidence of the classification. The classifier wrongly classifies all the generated images with 99\% confidence. In Figure 2, we report the confidence scores of the generated images. After iteration K, all the generated adversarial images trick the network successfully.  

In Table I, we report the image quality assessments of the generated examples. The quality of the images are higher, when regularization techniques are employed. 

\begin{table}[h]
\centering
\begin{tabular}{lll}
\hline
Metrics                                    & With & Without \\ \hline
Peak Signal to Noise Ratio \cite{b8}       & 370       & 350          \\ \hline
Universal Image Quality Index \cite{b9}    & 0.99      & 0.90         \\ \hline
Spatial Correlation Coefficient \cite{b10} & 0.85      & 0.21         \\ \hline
Visual Information Fidelity \cite{b11}     & 0.99      & 0.95         \\ \hline
\end{tabular}
\label{table:quality_imagenet}
\caption{First column shows the image quality metrics, the second and third  columns are the measures obtained  with regularization and  without regularization cost functions, respectively. For \cite{b8, b10, b11},  the maximum value is 1.0 and for \cite{b9}, the higher, the better. We report the average values of the metrics calculated over the 10 generated examples in ImageNet dataset.}
\vspace{-0.2cm}
\end{table}

\begin{figure}[h]

    \centering
    
    \begin{minipage}{0.48\textwidth}
    
        \centering
        \includegraphics[width=\textwidth]{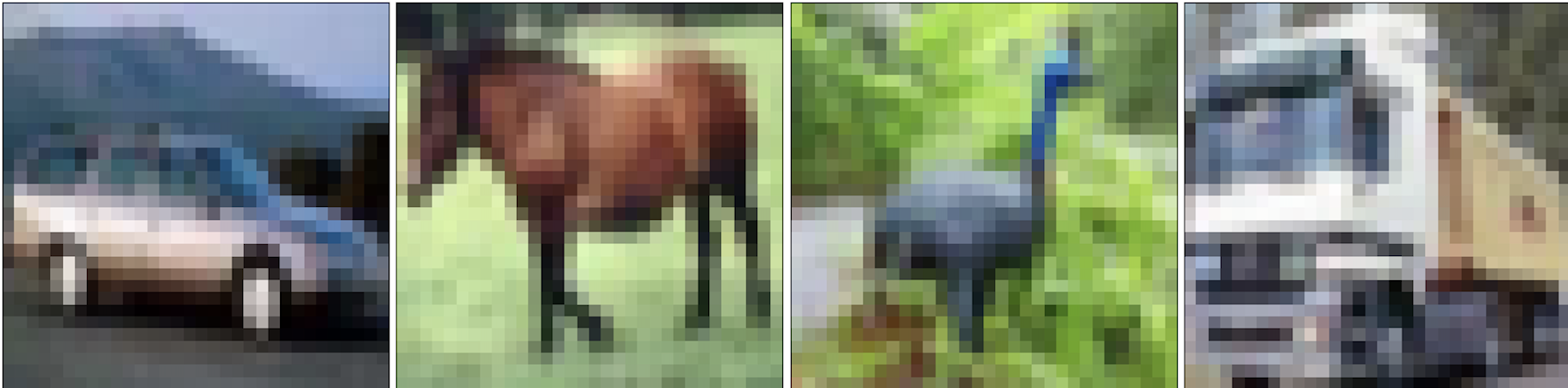} 
    \end{minipage}\hfill
    \begin{minipage}{0.48\textwidth}
        \centering
        \includegraphics[width=\textwidth]{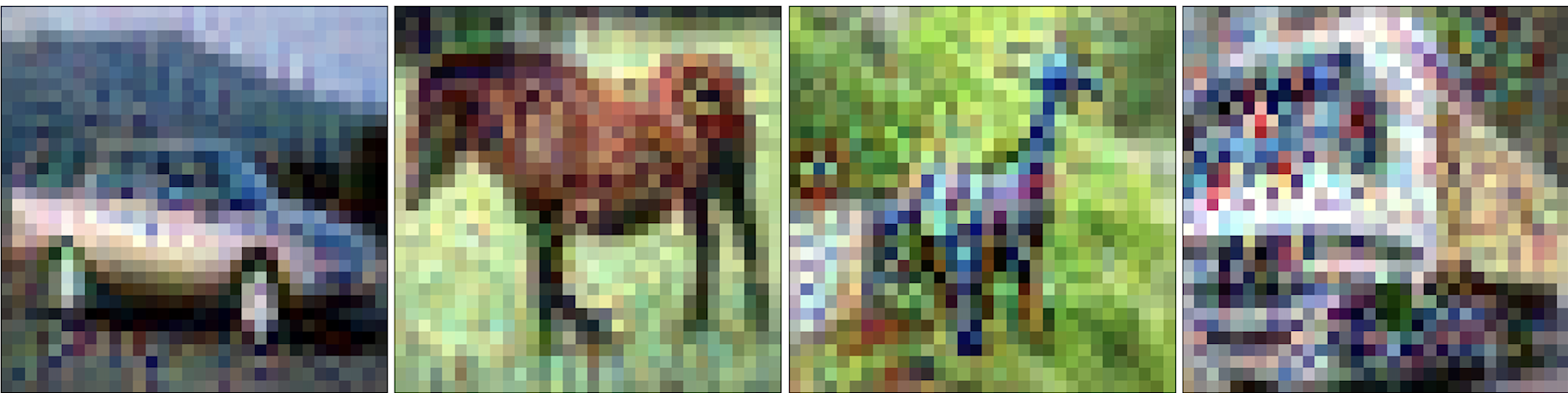} 
    \end{minipage}\hfill
    \begin{minipage}{0.48\textwidth}
        \centering
        \includegraphics[width=\textwidth]{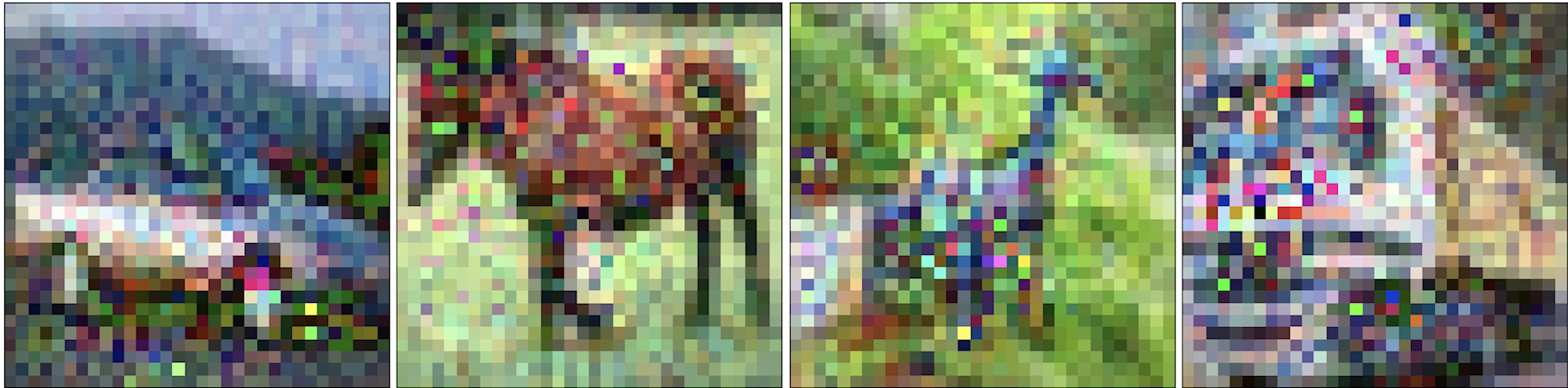} 
        \caption{Experiments in CIFAR10 dataset: Top row: Original images. Middle row: JND adversarial images, generated with regularization. Bottom row: Adversarial images generated without any regularization. All of the examples successfully trick the classifier.}
    \end{minipage}
\label{fig:cifar10_images}
\vspace{-0.45cm}
\end{figure}

\subsection{Classification Experiments on CIFAR10 Dataset}

We generated 8 different adversarial images, half with the regularization techniques and the other half without regularization techniques. All of the generated images successfully tricked the classifier. Using regularization does not significantly  affect the results, since the resolution of the images is $32x32$. The generated images can be seen in Figure 4. Visual comparison shows that  images generated with regularization are slightly better that the images generated without regularization.

We report the same image quality assessment metrics for the generated images in the Table II. The image quality scores are similar for generated images with and without using regularization. According to these results, we  can conclude that resolution of the images is an important parameter for generation of high quality adversarial images. 

\begin{table}[h]
\centering
\begin{tabular}{lll}
\hline
Metrics                                    & With  & Without  \\ \hline
Peak Signal to Noise Ratio \cite{b8}       &  341.14     & 341.21         \\ \hline
Universal Image Quality Index \cite{b9}    &  0.68   & 0.75       \\ \hline
Spatial Correlation Coefficient \cite{b10} & 0.06      & 0.06         \\ \hline
Visual Information Fidelity \cite{b11}     & 0.12      & 0.10         \\ \hline
\end{tabular}
\label{table:quality_cifar}
\caption{The left part is with regularization and the right part is without regularization. For \cite{b8, b10, b11},  the maximum value is 1.0 and for \cite{b9}, the higher, the better. We report the average values of the metrics calculated over the 10 generated examples in CIFAR10 dataset.}
\vspace{-0.5cm}
\end{table}

\subsection{Adversarial Image Generation for Object Detection}

For the object detection task, we used Retina-Net, which is pre-trained on Microsoft COCO \cite{b4}. We use a simple road image which can be seen in the Figure \ref{fig:obj_detect} and update it iteratively to generate an adversarial example, which tricks the classifier in a way that the classifier sees a car in the middle of the image. We have generated two different adversarial examples from the starting road image on the left in the Figure \ref{fig:obj_detect}. In the middle one, we did not use any regularization techniques. Therefore, the output adversarial examples has a noise which can be detected by  humans.  We also generate an adversarial example by adding regularization terms, which is capable of tricking the network. 

Both of the generated adversarial examples successfully trick the Retina-Net object detector. Although there is no car in the middle of the image, both examples generate a car at the output of the model. The confidences for the  middle  image is 89\%, whereas the confidence for the image in the right is 91\%.






\vspace{-0.3cm}

\section{Conclusion and Future Work}

In this study, we propose a new method for generating adversarial images which is almost indistinguishable by humans, yet can successfully trick a   machine learning algorithm based on Convolutional Neural Networks. For this purpose, we adopt the concept of Just Noticeable Difference (JND) of human perception to machine perception and we define just noticeably different adversarial images for models. Then, we formalize JND concept for machines by adding a set of regularization term into the cost function of the network. 

We used three different data sets, namely, Imagenet, CIFAR10 and MS COCO to train classification and object detection models. The generated JND adversarial images  can  successfully trick a pre-trained Inception v3 image classifier and a RetinaNet object detector.  We report the image quality metrics for generated JND images. According to our results, JND adversarial images look very authentic compared to the state of the art image generators, specifically when high resolution images are used.


In the future, Generative Adversarial Networks (GAN) can be used to generate adversarial images. JND adversarial images will be generated  by training a GAN in which an original image will be fed as an input to the generator and will be distorted gradually to trick the discriminator.


\end{document}